\documentclass[11pt,epsf]{article}
\usepackage{alltt}
\usepackage[all]{xy}
\usepackage{latexsym,amssymb,amsmath,amsfonts, amsthm, xcolor}

\usepackage{enumerate}
\usepackage{graphicx}
\graphicspath{ {./images/} }

\def\vbar{\mathchoice{\vrule height6.3ptdepth-.5ptwidth.8pt\kern-.8pt}
{\vrule height6.3ptdepth-.5ptwidth.8pt\kern-.8pt}
{\vrule height4.1ptdepth-.35ptwidth.6pt\kern-.6pt}
{\vrule height3.1ptdepth-.25ptwidth.5pt\kern-.5pt}}
\def\fudge{\mathchoice{}{}{\mkern.5mu}{\mkern.8mu}}
\def\bbc#1#2{{\rm \mkern#2mu\vbar\mkern-#2mu#1}}
\def\bbb#1{{\rm I\mkern-3.5mu #1}}
\def\bba#1#2{{\rm #1\mkern-#2mu\fudge #1}}
\def\bb#1{{\count4=`#1 \advance\count4by-64 \ifcase\count4\or\bba A{11.5}\or
\bbb B\or\bbc C{5}\or\bbb D\or\bbb E\or\bbb F \or\bbc G{5}\or\bbb H\or
\bbb I\or\bbc J{3}\or\bbb K\or\bbb L \or\bbb M\or\bbb N\or\bbc O{5} \or
\bbb P\or\bbc Q{5}\or\bbb R\or\bbc S{4.2}\or\bba T{10.5}\or\bbc U{5}\or
\bba V{12}\or\bba W{16.5}\or\bba X{11}\or\bba Y{11.7}\or\bba Z{7.5}\fi}}

\newcommand{\vs}{\vspace{0.25cm}}

\usepackage{amsmath}

\newtheorem{theorem}{Theorem}
\newtheorem{itlemma}{Lemma}[section]
\newtheorem{itproposition}[itlemma]{Proposition}
\newtheorem{itcorollary}[itlemma]{Corollary}
\newtheorem{itremark}[itlemma]{Remark}
\newtheorem{itremarks}[itlemma]{Remarks}
\newtheorem{itdefinition}[itlemma]{Definition}
\newtheorem{itexample}[itlemma]{Example}

\newenvironment{lemma}{\begin{itlemma}\rm}{\end{itlemma}} 
\newenvironment{remark}{\begin{itremark}\rm}{\end{itremark}} 
\newenvironment{remarks}{\begin{itremarks} \rm}{\end{itremarks}}
\newenvironment{corollary}{\begin{itcorollary}\rm}{\end{itcorollary}}
\newenvironment{proposition}{\begin{itproposition}\rm}{\end{itproposition}}
\newenvironment{definition}{\begin{itdefinition}\rm}{\end{itdefinition}}
\newenvironment{example}{\begin{itexample}\rm}{\end{itexample}}
\newenvironment{fact}{\noindent {{\bf Fact}}:\ \ }{\hfill \medskip}
\newenvironment{claim}{\noindent {\em Claim}. \ \ }{\hfill \medskip}
\newcommand{\be}[1]{\begin{equation}\label{#1}}
\newcommand{\ee}{\end{equation}}
\newcommand{\bl}[1]{\begin{lemma}\label{#1}}
\newcommand{\br}[1]{\begin{remark}\label{#1}}
\newcommand{\brs}[1]{\begin{remarks}\label{#1}}
\newcommand{\bt}[1]{\begin{theorem}\label{#1}}
\newcommand{\bd}[1]{\begin{definition}\label{#1}}
\newcommand{\bp}[1]{\begin{proposition}\label{#1}}
\newcommand{\bc}[1]{\begin{corollary}\label{#1}}
\newcommand{\bfact}[1]{\begin{fact}\label{#1}}
\newcommand{\bex}[1]{\begin{example}\label{#1}}
\newcommand{\ec}{\end{corollary}}
\newcommand{\efact}{\end{fact}}
\newcommand{\eex}{\end{example}}
\newcommand{\el}{\end{lemma}}
\newcommand{\er}{\end{remark}}
\newcommand{\ers}{\end{remarks}}
\newcommand{\et}{\end{theorem}}
\newcommand{\ed}{\end{definition}}
\newcommand{\ep}{\end{proposition}}
\newcommand{\epr}{\end{proof}}
\newcommand{\bpr}{\begin{proof}}
\newcommand{\bcl}{\begin{claim}}
\newcommand{\ecl}{\end{claim}}

\newcommand{\bi}{\begin{itemize}}
\newcommand{\ei}{\end{itemize}}
\newcommand{\ben}{\begin{enumerate}}
\newcommand{\een}{\end{enumerate}}

\usepackage{graphicx}

\begin{document}

\begin{center}{\Large
{\bf Algorithms for Quantum Control without Discontinuities; Application to the Simultaneous Control of two Qubits} }
\end{center}

\vs

\vs

\begin{center} 

Domenico D'Alessandro and Benjamin Sheller

Department of Mathematics, Iowa State University

{April 22, 2019}

\vs

\end{center}

\begin{abstract}

We propose a technique to design control algorithms for a class of finite dimensional quantum systems so  that  the control law does not present discontinuities. The class of models  considered admits a group of symmetries which allows us to reduce the  problem of control to
 a quotient space where the control system is `fully actuated'. As a result we can prescribe a desired trajectory which is, to some extent, arbitrary and derive the corresponding control. We discuss the application to the simultaneous control of two non-interacting spin $\frac{1}{2}$ particles with different gyromagnetic ratios in zero field NMR in detail. Our method provides a flexible toolbox for the design of control algorithms  to drive the state of finite dimensional quantum systems to any desired final configuration with smooth controls.

\end{abstract}

\vs

\vs

\vs


\vs

\vs

\vs 

\section{Introduction} The combination of control theory techniques with quantum mechanics (see, e.g., \cite{AltaTico}, \cite{Mikobook}, \cite{Optrev})  has generated a rich set  of control algorithms for quantum mechanical systems modeled by the 
Schr\"odinger (operator) equation 
\be{Scrod11}
\dot X=AX+ \sum_{j=1}^m B_j u_jX, \qquad X(0)={\bf 1}. 
\ee    
Here we assume that we have a finite dimensional model, $A$ and $ B_j$ are matrices in $\mathfrak{su}(n)$ for each $j=1,...,m$, $X$ is the unitary propagator, which is  equal to the identity ${\bf 1}$ at time zero, and $u_j$ are the controls. These are usually electromagnetic fields, constant in space but possibly time-varying, which are the output of an appropriately engineered  pulse-shaping device. Many of the proposed algorithms in the literature involve control functions which are only piecewise continuous and in fact have `jumps' at certain points of the interval of control. For example, control algorithms  based on {\it Lie group decompositions} (see, e.g., \cite{Rama1}) involve `switches' between different Hamiltonians; Algorithms based on 
{\it optimal control}, even if they produce smooth control functions, often require a jump at the beginning of the control interval in order for  the control to achieve the prescribed value in norm (assuming a bound in norm of the optimal control as in \cite{Bosca1}). Beside the practical  problem of generating (almost) instantaneous switches with pulse shapers, such discontinuities introduce undesired high frequency 
components in the dynamics of the controlled system. For these reasons, it is important  to have algorithms which produce {\it smooth} control functions whose values at the beginning and the end of the control interval are equal to zero. 

This paper describes a method to design control functions without discontinuities in order to drive the state of a class of quantum systems of the form (\ref{Scrod11}) to an arbitrary final configuration. Our main example of application will be the simultaneous control of two quantum bits in zero field NMR, a system which was also considered in \cite{Xinhua} in the context of optimal control. As compared to that paper, we abandon here the requirement of time optimality (under the requirement of bounded norm for the control) but introduce a novel 
method which will allow us more flexibility in the control design.  The result is a control algorithm that does not present discontinuities, with the control equal to zero at the beginning and at the end of the control interval.  

The paper is organized in two main sections each of which divided into several subsections. In Section \ref{GenTheor} we describe the class of systems we consider and the general theory underlying our method. We also present two simple examples of quantum systems where the theory applies. In Section \ref{App2QB} we detail the application to the system of two spin $\frac{1}{2}$ particles in zero field NMR above mentioned. This section includes a description of the model as well as the explicit numerical treatment of a control problem: the independent control of the two spin $\frac{1}{2}$ particles to two different types of Hadamard gates.

\section{General Theory}\label{GenTheor}

\subsection{Class of systems considered}\label{Classo} 

Consider the class of control systems (\ref{Scrod11}) with $A$ and $ B_j$, $j=1,...,m$, in $\mathfrak{su}(n)$ and let ${\cal L}$ denote the Lie algebra generated by $\{A, B_1,...,B_m\}$. We assume that ${\cal L}$ is semi-simple, which implies, since ${\cal L} \subseteq \mathfrak{su}(n)$, that the associated Lie group $e^{\cal L}$ is compact. The Lie algebra ${\cal L}$ is called, in quantum control theory, the {\it dynamical Lie algebra} associated to the system (\ref{Scrod11}).  Since $e^{\cal L}$ is compact, the Lie group $e^{\cal L}$ is the set of states for (\ref{Scrod11}) reachable by changing the control \cite{Suss}. In particular if ${\cal L}=\mathfrak{su}(n)$, the system is said to be {\it controllable} because every special unitary matrix can be obtained with appropriate control. These are known facts in quantum control theory (see, e.g., \cite{Mikobook}). We assume that ${\cal L}$ has a (vector space) decomposition ${\cal L}={\cal K} \oplus {\cal P}$, such that $[{\cal K}, {\cal K}] \subseteq {\cal K}$, i.e., ${\cal K}$ is a {\it Lie subalgebra of ${\cal L}$}, which we also assume to be semisimple so that $e^{\cal K}$ is compact. Moreover $[{\cal K}, {\cal P}] \subseteq {\cal P}$. A special case is when in addition $[{\cal P}, {\cal P} ] \subseteq {\cal K}$ in which case the decomposition ${\cal L}={\cal K} \oplus {\cal P}$ defines a symmetric space of $e^{\cal L}$ \cite{Helgason}. We assume, in the model (\ref{Scrod11}), that such a decomposition exists so that $A \in {\cal K}$ and $\{B_1,...,B_m\}$ forms a basis for ${\cal P}$.

Under such circumstances, we can reduce ourselves to the case $A=0$ in (\ref{Scrod11}), i.e., to systems of the form 
\be{Scro}
\dot U= \sum_{k=1}^m \hat u_k B_k U, \qquad U(0)={\bf 1}.   
\ee   
To see this, assume that for any fixed interval $[0,t_f]$ and any desired final condition $ U_f$, we are able to find a control $\hat u_k$ steering the state $U$ in (\ref{Scro}) from the identity ${\bf 1}$ to $U_f$. Let $a_{kj}=a_{kj}(t)$, $k,j=1,...,m$ the coefficients forming an $m \times m$ orthogonal matrix, so that, for any $j=1,...,m$,  
\be{transforB}
e^{-At} B_j e^{At}=\sum_{k=1}^m a_{kj}(t) B_k. 
\ee 
Let $X_f$ be  the desired final condition for (\ref{Scrod11}) and $\hat u_k$ be the control steering the state $U$ of system (\ref{Scro}) from the identity ${\bf 1}$ to $e^{-At_f} X_f$, in time $t_f$. Then the control $u_j$ obtained inverting 
\be{tobeinverted}
\hat u_k(t):=\sum_{j=1}^m a_{kj}(t)u_j(t), 
\ee
steers the state $X$ of (\ref{Scrod11}) from the identity to $X_f$. This follows from the fact that, if $U=U(t)$ is the solution of (\ref{Scro}) with the control $\hat u_k$, and final condition $e^{-At_f} X_f$,  then $X=e^{At}U$ is a solution of   (\ref{Scrod11}), with the controls $u_j$ given by (\ref{tobeinverted}) and therefore  the final condition at $t_f$ is $X_f$. Notice that the transformation (\ref{tobeinverted}) does not modify the smoothness properties of the control, neither does it modify the fact that the control is zero at the beginning and at the end of the control interval (or at any other point). 
Therefore in the following we shall deal with {\it driftless systems} of the form (\ref{Scro}) with the Lie algebraic ${\cal L}={\cal K} \oplus {\cal P}$, structure above described. In particular $\{B_1,...,B_m\}$ forms a basis for ${\cal P}$.  

\subsection{Symmetry reduction}
 
The compact Lie group $e^{\cal K}$ can be seen as a Lie 
transformation group which acts on $e^{\cal L}$  via conjugation $X \in e^{\cal L}  \rightarrow KXK^{-1}$, $ K \in e^{\cal K}$. Moreover this is  a {\it group of symmetries} for system (\ref{Scro}) 
in the sense that  $K B_j K^{-1} \in {\cal P}$ for each $j$, for every $K \in e^{\cal K}$. In particular let $K B_j K^{-1} :=\sum_{k=1}^m a_{kj} B_k$ for an orthogonal matrix $\{a_{kj}\}$ depending on $K \in e^{\cal K}$ (cf. (\ref{transforB})).  If $U=U(t)$ is a trajectory corresponding to a control $\hat u_k$, $KUK^{-1}$ is the trajectory 
corresponding to controls $u_k:=\sum_{j=1}^ma_{kj} \hat u_j$, as it is easily seen from (\ref{Scro}) and 
$$
K\dot U K^{-1}=\sum_{j=1}^m \hat u_j KB_j K^{-1} KUK^{-1}= \sum_{j=1}^m \hat u_j \left( \sum_{k=1}^m a_{kj} B_k \right) KUK^{-1}=$$
$$=\sum_{k=1}^m \left( \sum_{j=1}^m a_{kj} \hat u_j \right) B_k KUK^{-1}=\sum_{k=1}^m u_k B_k KUK^{-1}. 
$$  
This suggests to treat the control problem on the {\it quotient space} $e^{\cal L}/e^{\cal K}$ corresponding 
to the above action of $e^{\cal K}$ on $e^{\cal L}$.

From the theory of Lie transformation groups (see, e.g., \cite{Bredon}) we know that the quotient space $e^{\cal L}/e^{\cal K}$ has the structure of a {\it stratified space} where each stratum corresponds to an {\it orbit type}, i.e., a set of points in $e^{\cal L}$ which have conjugate isotropy groups. The stratum corresponding to the smallest possible isotropy group, $K_{min}$, is known  to be a connected manifold which is {\it open and dense} in $e^{\cal L}/e^{\cal K}$. We denote it here by $e^{\cal L}_{reg}/e^{\cal K}$, where $reg$ stands for the {\it regular} part.  Its preimage in $e^{\cal L}$, $e^{\cal L}_{reg}$, under the natural projection $\pi \, : \, e^{\cal L} \rightarrow  e^{\cal L}/e^{\cal K}$ is open and dense in $e^{\cal L}$ 
as well. This is called the {\it regular part} of   $e^{\cal L}/e^{\cal K}$, (resp. $e^{\cal L}$). The complementary set in $e^{\cal L}/e^{\cal K}$, (resp. $e^{\cal L}$) is called the {\it singular part}. The dimension of $e^{\cal L}_{reg}/e^{\cal K}$ as a manifold is 
\be{dimensio}
\dim (e^{\cal L}_{reg}/e^{\cal K}) =\dim (e^{\cal L})-\dim (e^{\cal K})+\dim K_{min}=\dim ({\cal L})-\dim ({\cal K})+(\dim K_{min}), 
\ee   
where $\dim ( K_{min})$ is the dimension of the minimal isotropy group as a Lie group.\footnote{More discussion on these basic facts in the theory of Lie transformation groups can be found in \cite{NOIJDCS} and references therein.}   In particular, if 
$K_{min}$ is a {\it discrete  Lie group}, i.e., it has dimension zero, the right hand side of (\ref{dimensio}) is the dimension of the subspace ${\cal P}$.  This is verified  for instance in $K-P$ problems (cf., e.g., \cite{conBenECC})  when $e^{\cal L}=SU(n)$. We shall assume this to be the case in the following. 

According to a result in \cite{conBenECC}, under the assumption that the minimal isotropy group $K_{min}$ is discrete,   the restriction of $\pi_*$ to $R_{x*} {\cal P}$ is an isomorphism onto $T_{\pi(x)} \left( e^{\cal L}_{reg}/{e^{\cal K}} \right)$ for each point $x$ in the regular part, $e^{\cal L}_{reg}$.  Here, as it is often done, we have identified the Lie algebra ${\cal L}$ with the tangent space of $e^{\cal L}$ at the identity ${\bf 1}$, and therefore ${\cal P}$ is  identified with a subspace of the tangent space at $\bf{1}$. The map $R_x$ denotes the {\it right translation} by $x$ so that $R_{x*}{\cal P}$ is a subspace (with the same dimension) of the tangent space at $x$, $T_x {e^{\cal L}}$.\footnote{Recall that for a map $f \, : \, M \rightarrow N$ for two manifolds $M$ and $N$, $f_*$ denotes the {\it differential} (also called {\it push-forward}) $f_* \, : \, T_xM \rightarrow  T_{f(x)}N$ between two tangent spaces. When we want to emphasize the point $x$ we write $f_*|_x$.  } In Appendix B, we show that in given coordinates the determinant of the restriction of  $\pi_*$ to $R_{x*}{\cal P}$ is invariant under the action of $e^{\cal K}$. The above isomorphism result says that in the regular part $\det \pi_* \not=0$. In this situation, for every regular point $U \in e^{\cal L}$, for every tangent vector $V \in  T_{\pi(U)}(e^{\cal L}_{reg}/e^{\cal K})$, we can find a tangent vector $\pi_*^{-1} V \in R_{U*} {\cal P}$. Such a tangent vector is {\it horizontal} for system (\ref{Scro})  which means  that it  can be written as a linear combination of the available vector fields $\{  B_k U \}$ in (\ref{Scro}). If $\Gamma=\Gamma(t)$ is a curve entirely contained in $e^{\cal L}_{reg}/e^{\cal K}$ and $U=U(t)$ a curve in $e^{\cal L}_{reg}$ such that $\pi(U(t))=\Gamma(t)$ for every $t$, i.e., $U$ is a `{\it lift}' of $\Gamma$, then $\pi_*|_{U}^{-1} \dot \Gamma$ is a horizontal tangent vector at $U(t)$ for every $t$. If $\Gamma$ joins two points $\Gamma_0$ and $\Gamma_1$ in  $e^{\cal L}_{reg}/e^{\cal K}$, in the interval $[t_0,t_1]$,  and $U_0$ is such that $\pi(U_0)=\Gamma_0$, then the solution of the differential system 
\be{diffsys}
\dot U=\pi_*|_{U}^{-1} \dot \Gamma, \qquad U(t_0)=U_0,  
\ee 
is such that $\pi(U(t_1))=\Gamma_1$. Therefore, once we prescribe an arbitrary trajectory $\Gamma$ to move in the quotient space between two given orbits $\Gamma_0$ and $\Gamma_1$  in the regular part,  the control specified by 
\be{deficontr}
\pi_*|_{U}^{-1} \dot \Gamma=\sum_{j=1}^m u_j B_j U
\ee
 will allow us to move between two states $U_0$ and $U_1$ such that $\pi(U_0)=\Gamma_0$ and $\pi(U_1)=\Gamma_1$.

 \subsection{Methodology for Control} 
 The above treatment suggests a general methodology to design control laws 
  for  systems of the form (\ref{Scro}) described in subsection \ref{Classo}. In fact, 
   given the freedom in the choice of the trajectory $\Gamma=\Gamma(t)$ above mentioned,  
   we can design such controls satisfying various requirements and in particular  without discontinuity. Such a  methodology can be summarized as follows. 
   
 First of all we need to obtain a geometric description of the orbit space $e^{\cal L}/e^{\cal K}$, and in particular of its regular part  $e^{\cal L}_{reg}/e^{\cal K}$, and verify that the minimal isotropy group, which is the isotropy group of the elements in $e^{\cal L}_{reg}$, is discrete so that the right hand side of 
 (\ref{dimensio}) is equal to $\dim {\cal P}$. This is a weak assumption, easily verified in the examples that will follow and that can be proven true in several cases \cite{conBenECC}, \cite{BenTesi}. Then one chooses
  coordinates for the manifold $e^{\cal L}_{reg}/e^{\cal K}$. These are   expressed in terms of the original coordinates in 
 $e^{\cal L}$ or, more commonly, in terms of the entries of the matrices in $e^{\cal L}$. Such coordinates are a complete set of {\it independent invariants} with respect to the (conjugacy) action of the group $e^{\cal K}$. The word `complete' here means that the knowledge of their values uniquely determines the {\it orbit}, i.e., a point in $e^{\cal L}_{reg}/e^{\cal K}$. There are $m=\dim({\cal P})$ of them, as this is the dimension of $e^{\cal L}_{reg}/e^{\cal K}$ (cf. (\ref{dimensio})).  Once we have coordinates $\{x^1,...,x^m\}$, the tangent vectors $\left \{ \frac{\partial}{\partial x^1},...,\frac{\partial}{\partial x^m}\right\}$ at every regular point in the quotient space determine a basis of the tangent space of  $e^{\cal L}_{reg}/e^{\cal K}$. For any trajectory $\Gamma$ in 
 $e^{\cal L}_{reg}/e^{\cal K}$,
  we can write the tangent vector $\dot \Gamma(t)$ as  $\dot \Gamma(t)=\sum_{j=1}^m \dot x^{j}\frac{\partial}{\partial x^j}$, for some functions $\dot x^j$. With this choice of coordinates, one then needs to calculate, for every regular point $U$,  the  matrix for $\pi_*|_U$ as restricted to $R_{U*} {\cal P}$  and its inverse $\pi_*|_U^{-1}$. This allows us to implement 
 formula (\ref{deficontr}) to obtain the control from a given prescribed trajectory in the orbit space.

 We remark that there is an issue concerning the fact that our initial condition which is the identity ${\bf 1}$ in (\ref{Scro}) (and possibly the final desired condition)  is  not in the regular part of $e^{\cal L}$.  In fact the whole Lie group $e^{\cal K}$ is the isotropy group of the identity. If we take for $\Gamma$ a trajectory which starts from the class corresponding to the identity, the matrix corresponding to $\pi_*$ may become  singular as $t \rightarrow 0$ and therefore it will be impossible to derive the control directly from formula (\ref{deficontr}). This problem can be overcome by applying a {\it preliminary control} which takes the state of system (\ref{Scro}) out of the singular part and into  the regular part of $e^{\cal L}$.  To avoid discontinuities, such a control is chosen to be zero at the beginning and at the end of the control interval. It takes the system to a point $U_0$ with $\pi(U_0)=\Gamma_0 \in e^{\cal L}_{reg}/e^{\cal K}$. Then we choose the trajectory in the quotient space $\Gamma$ in the regular part of the quotient space which joins $\Gamma_0$ and $\Gamma_1$ where $\Gamma_1$ is the orbit of the desired final condition. The control obtained through (\ref{deficontr}) will steer system (\ref{Scro}) to a state $\hat U_f$ in the same orbit as the desired final condition. Therefore if $U_f$ is the desired final condition we will have 
 $\pi(\hat U_f)=\pi(U_f)=\Gamma_1$.  Notice that we also want $\dot \Gamma\to 0$ at both the initial and final point so that the control is zero and can be {\it concatenated} continuously with the preliminary control above described.

 It is possible that the desired final condition $U_f$ is also in the singular part of $e^{\cal L}$. This problem can be tackled in two ways. We can recall that the regular part is open and dense and therefore we can always drive to a state in the regular part arbitrarily close to the desired $U_f$. This means that our algorithm  will only give an approximate control, but which will steer the system arbitrarily close to $U_f$. Alternatively we can select a regular element $S \in e^{\cal L}$ and such that $U_f S^{-1}$ is also regular.\footnote{Such an element $S\in e^{\mathcal{L}}_{\text{reg}}$ always exists for any $U\in e^{\cal L}$, by the following argument: Assume that it does not exist. Then for every regular $S$, $US$ is singular. Therefore, by indicating by $L_U$ the left translation by $U$ we have,  $L_U(e^{\mathcal{L}}_{\text{reg}})\subseteq e^{\mathcal{L}}_{\text{sing}}$. Then by applying the unique bi-invariant Haar measure $\mu$ on $e^{\cal L}$ with $\mu(e^{\cal L})=1$ implies $\mu(e^{\mathcal{L}}_{\text{reg}})=\mu(L_U(e^{\mathcal{L}}_{\text{reg}}))\leq\mu(e^{\mathcal{L}}_{\text{sing}})$. On the other hand, $\mu(e^{\mathcal{L}}_{\text{sing}})=0$ since $\mu$ must also correspond to the Riemannian volume of the bi-invariant Killing metric (normalized if necessary) and each stratum in $e^{\mathcal{L}}_{\text{sing}}$ has dimension strictly less than dimension of $e^{\cal L}$ and thus has volume $0$ and therefore invariant measure $0$. But $\mu(e^{\mathcal{L}}_{\text{reg}})=\mu(e^{\mathcal{L}})-\mu(e^{\mathcal{L}}_{\text{sing}})=\mu(e^{\mathcal{L}})=1$. This is a contradiction.} Then we find two controls: $u_1$ driving $U$ in (\ref{Scro}) from the identity to $S$ in (\ref{Scro}) and $u_2$ driving $U$ in (\ref{Scro}) from the identity to $U_fS^{-1}$ in (\ref{Scro}). In particular, because of the right invariance of system (\ref{Scro}), $u_2$ also drives $S$ to $U_f$. 
 Therefore, the concatenation of $u_1$ (first) and $u_2$ (second) will drive to the desired final configuration. Therefore in the following, for simplicity, we shall assume that the final desired state is in the regular part.

 The (concatenated) control $\hat u_j$  obtained from the tangent vector $\dot \Gamma$ at every time $t$ for a trajectory on the quotient space $\Gamma$ (cf. (\ref{deficontr})) will drive the state $U$ of (\ref{Scro}) from the identity ${\bf 1}$ only to a state $\hat U_f$ which is in the same orbit as the desired final state $U_f$. 
  There exists $K \in e^{\cal K}$ such 
 that $K\hat U_f K^{-1}=U_f$. Once such a  $K$ is found it will determine through 
 $\sum_{j=1}^m KB_j \hat u_jK^{-1}=  \sum_{k=1}^m B_k  u_k$ the actual control $\{u_k\}$ to apply. We remark that this tranformation does not modify the smoothness properties of the control, nor the fact that it is zero at some point (in particular at the beginning and at the end of the control interval).

\subsection{Examples}
 
\subsubsection{Control of a single  spin $\frac{1}{2}$ particle}\label{Twol}

Consider the Schr\"odinger operator equation (\ref{Scro}) in the form 
\be{Scro1}
\dot U=\begin{pmatrix} 0 & \alpha(t) \cr 
-\alpha^*(t) & 0
\end{pmatrix} U, \qquad U(0)={\bf 1}_2, 
\ee
with $U$ in $SU(2)$. The complex-valued function $\alpha$ is a 
complex control field representing the $x$ and $y$ 
components of an electro-magnetic field. 
The dynamical Lie algebra ${\cal L}$ is $\mathfrak{su}(2)$ and it has a decomposition 
$\mathfrak{su}(2)={\cal K} \oplus {\cal P}$ with ${\cal K}$ diagonal and ${\cal P}$ 
anti-diagonal matrices.  
 The one-dimensional 
Lie group of {\it diagonal} matrices in $SU(2)=e^{\cal L}$ is a symmetry group $e^{\cal K}$ 
for the above system and the structure of the quotient space $SU(2)/e^{\cal K}$ is that of 
a closed unit disc \cite{NoiAutomat}. The entry $u_{1,1}$ of 
$U \in SU(2)$, which is a complex number with absolute value $\leq 1$, 
determines the orbit of the matrix $U$. The regular part 
of $SU(2)$ corresponds to  matrices with $|u_{1,1}| < 1$, i.e., the interior 
of the unit disc $\mathring D$, in the complex plane. The singular part is 
the boundary of the unit disc. Denote by $z$ the (complex) coordinate in the 
interior of the complex unit disc. This corresponds  to two {\it real} coordinates invariant under the action of $e^{\cal K}$ 
 (conjugation by diagonal matrices). Let $\Gamma=\Gamma(t)$ be a desired trajectory inside 
 the unit disc, which we denote by $z_d$ in the chosen coordinates.  The tangent 
vector $\dot \Gamma$ in (\ref{diffsys}) is given in complex coordinates by $\dot{z_d} \frac{\partial}{\partial z}$.\footnote{This means $\dot x_d \frac{\partial}{\partial x_d} +\dot y_d \frac{\partial}{\partial y_d}$ where $x_d$ and $y_d$ are the real and imaginary parts of $z_d$.} In the coordinates on $SU(2)$ used in equation (\ref{Scro}) the corresponding tangent vector for $\dot U$ is given by (cf. (\ref{Scro1})) $\begin{pmatrix} 0 & \alpha \cr -\alpha & 0 \end{pmatrix} U$ and the value of the control $\alpha$ is obtained by solving (\ref{diffsys}) which gives 
\be{diffebis}
\dot z_d=\frac{d}{dt}|_{t=0} z \left( e^{\begin{pmatrix}0 & \alpha \cr -\alpha^* & 0  \end{pmatrix}t} U \right), 
\ee
where $z(P)$ denotes the the $(1,1)$ entry of the matrix $P$. Equation (\ref{diffebis}) gives $\alpha=\frac{\dot z_d}{u_{21}}$, which,  
as expected from the above recalled  isomorphism theorem of \cite{conBenECC},  gives a one-to-one correspondence  between $\alpha$ and $\dot z_d$ as long as $U$ is in the 
regular part of $SU(2)$, i.e., it is not diagonal, i.e., $u_{2,1} \not=0$.


\subsection{Control of a three level system in the $\Lambda$ configuration}\label{Lambda}
Consider a three level quantum system where the controls couple level $|1\rangle $ to level $|2\rangle $ and level $|1 \rangle $ to level $|3\rangle$ but not level $|2\rangle$ and $|3\rangle$ directly. Assuming that $|1\rangle$ is the highest energy level, the energy level diagram takes the so-called $\Lambda$ configuration (see, e.g., \cite{Lambda1}). The Schr\"odinger operator equation (\ref{Scro}) is such that 
\be{lambdascro}
\sum_{j=1}^m u_j B_j =\sum_{j=1}^4 u_j B_j=
\begin{pmatrix}0 & \alpha & \beta \cr -\alpha^* & 0 & 0 \cr 
- \beta^* & 0 & 0 \end{pmatrix},  
\ee
with the complex control functions  $\alpha$ and $\beta$. Such a system admits a group of symmetries given by $e^{\cal K}=S(U(1) \times U(2))$, i.e., block diagonal matrices in $SU(3)$ with one block of dimension $1 \times 1$ and one block of dimension $2 \times 2$, and determinant equal to $1$. The Lie subalgebra ${\cal K}$ consists of matrices in $\mathfrak{su}(3)$ with a block diagonal structure with one block of dimension $1 \times 1$ and one block of dimension $2 \times 2$. The complementary subspace ${\cal P}$ is spanned by 
antidiagonal matrices in $\mathfrak{su}(2)$ with the same partition of rows and columns.   Such a system  was studied in \cite{ADS} in the context of optimal control and a description of the orbit space $SU(3)/e^{\cal K}$ was given. 
It was shown  that the regular part $SU(3)_{reg}/e^{\cal K}$ is 
homeomorphic to the product of two open unit discs 
$\mathring D_1 \times \mathring D_2$ in the complex plane. Up to a similarity transformation in  $e^{\cal K}=S(U(1) \times U(2))$, a matrix $U$ in $SU(3)$ 
can be written as   
\be{canonicalformSU3}
U=\begin{pmatrix} z_1 & \sqrt{1-|z_1|^2} & 0 \cr 
-\sqrt{1-|z_1|^2} & z_1^* & 0 \cr 0 & 0 & 1 \end{pmatrix}
\begin{pmatrix} 1 & 0 & 0 \cr 0 & z_2 & w \cr 0 & - w^* & z_2^* \end{pmatrix},   
\ee
for complex numbers $z_1,$ $z_2$ and $w$,  where $|z_1|\leq 1$ and $|z_2| \leq 1$. Strict inequalities  hold  if and only if $U$ is in the regular part in which case $z_1$ and $z_2$ can be taken as the coordinates in $SU(3)_{reg}/e^{\cal K}$. An alternative set of (complex) coordinates is given by $(z_1,T)$  where $T$ is the trace of the (lower) $2 \times 2$ block of the element $U \in SU(3)$ which is invariant (along with $z_1$) under the conjugation  action of elements in $e^{\cal K}=S(U(1) \times U(2))$. The  coordinates $(z_1,T)$ are related to the coordinates $(z_1,z_2)$ by  (from (\ref{canonicalformSU3})) $T=z_1^*z_2+z_2^*$ which is inverted as $z_2=\frac{T^*-z_1 T}{1-|z_1|^2}$. A desired trajectory $\Gamma$ in $SU(3)_{reg}/S(U(1)\times U(2))$ is written in these coordinates as $(z_{1d},T_d):=  (z_{1d}(t),T_d(t))$ . The associated tangent vector $\dot \Gamma$ in (\ref{diffsys})  is 
 $\dot z_{1d}(t) \frac{\partial}{\partial z_{1d}}+\dot T_{d} \frac{\partial}{\partial T}$. By applying $\pi_*|_{U}\dot U$ in (\ref{diffsys}) to $z_1$ and $T$ with the restriction  that $\dot U$ is of the form   $\begin{pmatrix}0 & \alpha & \beta \cr -\alpha^* & 0 & 0 \cr 
- \beta^* & 0 & 0 \end{pmatrix}U$, we obtain two equations for $\alpha$ and $\beta$, 
$$
\alpha u_{2,1}+\beta u_{3,1}=\dot z_{1d}, \qquad -\alpha u_{1,2}^*-\beta u_{1,3}^*=\dot T_{d}^*.  
$$ 
These are solved, using $\hat D:=u_{1,3}^*u_{2,1}-u_{3,1} u_{1,2}^*$ by 
\be{alphabetasol}
\alpha=\frac{u_{1,3}^* \dot z_{1,d}+ u_{3,1} \dot T_d^*}{\hat D}, \qquad 
\beta=-\frac{u_{1,2}^* \dot z_{1,d}+ u_{2,1} \dot T_d^*}{\hat D}. 
\ee

The  quantity $\hat D$ is different from zero if and only if the  matrix $U$ is in the regular part of $SU(3)$. This can be shown  in two steps: First one shows that $\hat D$ is invariant under the action of $S(U(1) \times U(2))$ by writing a matrix in  
$S(U(1) \times U(2))$ with an Euler-type decomposition 
as $F_1 R F_2$ with $F_1$ and $F_2$ diagonal and $R$ of the form 
$
R=\begin{pmatrix} 1 & 0 & 0 \cr 
0 & \cos(\theta) & \sin (\theta) \cr 
0 & -\sin(\theta) & \cos(\theta) 
\end{pmatrix}, 
$
and verifying that conjugation by each factor in $F_1 R F_2$  leaves $\hat D$ unchanged. The second step is to verify that $\hat D$ for the matrix $X$ in the form 
(\ref{canonicalformSU3}) is different from zero if and only if $|z_1| \not=1$ and 
$|z_2| \not=1$. This gives a quick  
test to check whether a matrix is in the regular part, i.e., if its isotropy group is the smallest possible one, which, in this case,  is the group of scalar matrices in $SU(3)$. 
{This fact also follows from the result in Appendix B which shows in general that $\det(\pi_*)$, with $\pi_* :\, R_{p *}{\cal P} \rightarrow T_{\pi(p)} e^{\cal L}_{reg}/e^{\cal K}$ is  invariant under the action of $e^{\cal K}$. }

As always, we have the problem that the initial point ${\bf 1}$ is  in the singular part of the orbit space decomposition and therefore $\hat D$ in (\ref{alphabetasol}) is zero at time $0$. As suggested in the previous section, we can however apply a preliminary control to 
steer away from the singular part.

\section{Simultaneous control of two independent spin $\frac{1}{2}$ particles}\label{App2QB}

\subsection{The model}
The dynamics of two spin $\frac{1}{2}$ particles with different gyromagnetic ratios in zero field NMR can be described by the Schr\"odinger equation (\ref{Scro}) (after appropriate normalization) where 
\be{Hamilto}
\sum_{k=1}^m \hat u_k B_k:=\sum_{x,y,z} u_{x,y,z}(t)( i\sigma_{x,y,z} \otimes {\bf 1}+ \gamma
 i {\bf 1} \otimes \sigma_{x,y,z}).   
\ee  
Here $u_{x,y,z}$ are the controls representing the $x,y,z$ components of the electromagnetic field, and $\sigma_{x,y,z}$ are the Pauli matrices defined as 
\be{Pauli}
\sigma_x:=  \begin{pmatrix} 0 & 1 \cr 1 & 0 \end{pmatrix}, \qquad
\sigma_y:=  \begin{pmatrix} 0 & -i \cr i & 0 \end{pmatrix}, \qquad
\sigma_z:=  \begin{pmatrix} 1 & 0 \cr 0 & -1 \end{pmatrix}.  \qquad
\ee
The parameter $\gamma$ is the ratio of the two gyromagnetic ratios and 
we shall assume that  $|\gamma|\not=1$. Under this assumption,  the dynamical Lie algebra ${\cal  L}$ for system (\ref{Scro}), (\ref{Hamilto}) is the $6-$dimensional Lie algebra spanned by $\{\sigma_1 \otimes {\bf 1} + {\bf 1} \otimes \sigma_2 \, | \, \sigma_1, \sigma_2 \in \mathfrak{su}(2)\}$.\footnote{This Lie algebra is isomorphic to $\mathfrak{so}(4)$.}   The corresponding Lie group $e^{\cal L}$, which is the set of reachable states for system  (\ref{Scro}), (\ref{Hamilto}), is $\{X_1 \otimes X_2 \, | \, X_1, X_2 \in SU(2)\}$, i.e. the tensor product $SU(2)\otimes SU(2)$. It is convenient to slightly relax the description of the state space and look at system (\ref{Scro}), (\ref{Hamilto}) as a system on the Lie group $SU(2) \times SU(2)$, i.e., the Cartesian direct product of $SU(2)$ with itself, and the dynamical equations (\ref{Scro}), (\ref{Hamilto}) 
replaced by
\be{reple}
\dot U= \sigma(t)U, \quad U(0)={\bf 1}, \qquad \dot V= \gamma \sigma(t)V, \quad V(0)={\bf 1},  
\ee 
with $\sigma(t):=\sum_{x,y,z} i u_{x,y,z}(t) \sigma_{x,y,z}$. 
The controls that drive system (\ref{reple}) to $(\pm U_f, \pm V_f)$ drive system (\ref{Scro}), (\ref{Hamilto}) to the state $U_f \otimes V_f$. Therefore we shall focus on the steering problem for system (\ref{reple}) which consists of steering one spin to $U_f$ and the other to $V_f$, simultaneously. Since $|\gamma| \not=1$, the dynamical Lie algebra associated with (\ref{reple}) is spanned by the pairs $(\sigma_1, \sigma_2)$ with $\sigma_1$ and $\sigma_2$ in $\mathfrak{su}(2)$. Such a Lie algebra can be written as ${\cal K} \oplus {\cal P}$ with ${\cal K}$ spanned by elements  of the form 
$(\sigma, \sigma)$ with $\sigma \in \mathfrak{su}(2)$ and ${\cal P}$ spanned by elements of the 
form $(\sigma, \gamma \sigma)$ with $\sigma \in \mathfrak{su}(2)$. At every $p \in G=SU(2) \times SU(2)$, the vector fields in (\ref{reple}) belong to $R_{p*} {\cal P}$. 

\subsection{Symmetries and the the structure of the quotient space}  

The Lie group $SU(2)$ acts on $G:=SU(2) \times SU(2)$ by {\it simultaneous conjugation}, that is, for $K \in SU(2)$, $(U_f, V_f) \rightarrow (K U_f K^{-1}, K V_f K^{-1})$ and this 
is a group of symmetries for system (\ref{reple}) in that  if $\sigma=\sigma(t)$ is the control steering to $(U_f,V_f)$, then $K\sigma K^{-1}$ is the control steering to $(KU_f K^{-1}, KV_f K^{-1})$. The quotient space of 
$SU(2) \times SU(2)$ under this action, $(SU(2) \times SU(2))/SU(2)$ was described in \cite{Xinhua} as follows. 

Consider 
a pair $(U_f, V_f)$ and let $\phi \in [0,\pi]$ be a real number so that the two eigenvalues of $U_f$ are $e^{i\phi}$ 
and $e^{-i\phi}$. If $0< \phi < \pi$ then $U_f \not= \pm {\bf 1}$ and there exists a unitary 
matrix $S$ such that $SU_f S^{\dagger}:=D_f$ is diagonal. Therefore the matrix $(U_f, V_f)$ is in the same orbit as $(D_f, SV_f S^\dagger)$. The parameter $\phi$ determines the orbit, along with the $(1,1)-$entry of $SV_f S^\dagger$, which does not depend on the choice of $S$.\footnote{All the possible diagonalizing matrices differ by a diagonal factor that does not affect the $(1,1)$ entry of  $SV_f S^\dagger$.} Such a $(1,1)$-entry has absolute value $\leq 1$ and therefore it is an element of the unit disk in the complex plane. The orbits corresponding to the values of $ 0< \phi < \pi$ (for the eigenvalue of the first matrix) are in one-to-one correspondence with the points  of a solid cylinder with height  equal to $\phi$. When $\phi=0$ (or $\phi=\pi$), the matrix $U_f$ is $\pm$ identity  and therefore the equivalence class  is determined by the eigenvalue of the matrices $V_f$, which are $e^{\pm i \psi}$ for $\psi\in [0, \pi]$. In the geometric description, the solid cylinder degenerates at the two ends to become a segment $[0,\pi]$. The regular part of the orbit space $G_{reg}$ is represented by points in the interior of the solid cylinder. Such points 
correspond to pairs 
$$
\left( \begin{pmatrix} e^{i \phi} & 0 \cr 0 & e^{-i\phi} \end{pmatrix},  \begin{pmatrix} z & w \cr -w^* & z^* \end{pmatrix}\right)
$$ 
with $\phi \in (0,\pi)$ and $|z|<1$. For these pairs,  the isotropy group is the discrete group $\{ \pm {\bf 1} \}$. In general points that are in the singular part correspond to pairs of matrices $(U_f, V_f)$ which can be simultaneously diagonalized. Therefore 
the condition that they commute 
\be{commut34}
U_f V_f=V_fU_f, 
\ee 
is necessary and sufficient for a pair $(U_f,V_f)$ to belong to the singular part.

Assume that $p$ is a regular point in $G_{reg}$ for this problem and $\pi$ is the natural projection $\pi: G_{reg} \rightarrow G_{reg}/SU(2)$.  Then from the theory in the previous section, the differential $\pi_*|_{p}$ is an isomorphism from $R_{p*}{\cal P}$ to $T_{\pi(p)} G_{reg}/SU(2)$. Let us choose a basis for ${\cal P}$ given by $(i\sigma_{x,y,z}, \gamma i\sigma_{x,y,z})$. To choose the three coordinates in $G_{reg}/SU(2)$, we consider a general element $p$ in $SU(2) \times SU(2)$ written as 
\be{puntop}
p:=(U_f,V_f):=\left( \begin{pmatrix} x & y \cr -y^* & x^* \end{pmatrix}, \begin{pmatrix} z & w \cr 
- w^* & z^* \end{pmatrix}\right). 
\ee 
For a complex number $q$ we shall denoted by $q_R:=\texttt{Re}(q)$ and $q_I:=\texttt{Im}(q)$. Notice that in (\ref{puntop}) we have 
$$
x_R^2+x_I^2+y_R^2+y_I^2=z_R^2+z_I^2+w_R^2+w_I^2=1. 
$$ 
Coordinates in $G_{reg}/SU(2)$ must be independent invariant functions of $(U_f,V_f)$ in (\ref{puntop}). We choose  
\be{ccordinates}
x^1:=x_R, \qquad x^2:=z_R, \qquad x^3:=x_I z_I+w_R y_R+ w_I y_I.
\ee
It is a direct verification to check that at any point $p \in SU(2) \times SU(2)$, $x^1$, $x^2$ and $x^3$ are unchanged by the (double conjugation) action of $SU(2)$, i.e., they are invariant. We remark also that we can consider two unit 
vectors $\vec V:=(x_R,x_I, y_R, y_I)$, and  $\vec W:=(z_R,z_I, w_R, w_I)$, and, if we do that, 
 $x^3=\vec V \cdot \vec W-x_Rz_R$. 

\subsection{Choice of invariants}
We pause a moment to detail how the invariant coordinates in (\ref{ccordinates})
 were chosen. We do this  because the method can be used for other examples. We consider the vectors $\vec V:=[x_R, x_I, y_R, y_I]^T$ and  
 $\vec W:=[z_R, z_I, w_R, w_I]^T$ and the adjoint action of $SU(2)$ on $SU(2) \times SU(2)$ which gives a linear action on $\vec Q:=[\vec V^T, \vec W^T]^T$. We are looking for functions $f=f(\vec V, \vec W)$ invariant under this action.  Given that every element of $SU(2)$ can be written according to Euler's decomposition as $e^{i \sigma_z \alpha} e^{i \sigma_y \theta}  e^{i \sigma_z \beta}$, for real parameters $\alpha, \beta$ and $\theta$, it is enough that $f$ is invariant  with respect to transformations of the form 
$e^{i \sigma_z \beta}$ and $e^{i \sigma_y \theta}$, for general real $\beta$ and $\theta$, in order for $f$ to be invariant with respect to all of $SU(2)$. If $X_z:=X_z(\beta):=e^{i \sigma_z \beta}$ then $Ad_{X_z}$ acting on  $[\vec V^T, \vec W^T]^T$ is 
\be{adXz}
Ad_{X_z(\beta)}:=\begin{pmatrix} 1 & 0  & 0 & 0 \cr 0 & 1 & 0 & 0 \cr 
0 & 0 & \cos(\beta) & - \sin (\beta) \cr 
0 & 0 & \sin(\beta) & \cos(\beta) 
\end{pmatrix}. 
\ee
If $X_y:=X_y(\theta):=e^{i \sigma_y \theta}$ then $Ad_{X_y}$ acting on  $[\vec V^T, \vec W^T]^T$ is 
\be{adXy}
Ad_{X_y(\theta)}:=\begin{pmatrix} 1 & 0  & 0 & 0 \cr 0 & \cos(\theta)  & 0 & \cos(\theta) \cr 
0  & 0  & 1 & 0 \cr 
0 & -\sin(\theta) & 0 & \cos(\theta) 
\end{pmatrix}. 
\ee
We first look for {\it linear invariants}, i.e., invariant functions $f$ of the form 
$f(\vec V, \vec W):=\vec a^{\, T} \vec V + \vec b^{\, T} \vec W$. From the condition
$$
\vec a^{\, T} \vec V + \vec b^{\, T} \vec W=\vec a^{\, T} Ad_{X}\vec V + \vec b^{\, T} Ad_{X}\vec W, 
$$ 
where $X$ may be equal to $X_z(\beta)$ or $X_{y}(\theta)$, with arbitrary $\beta$ and $\theta$, we find that the last three components of $\vec a$ and $\vec b$ must be zero. Therefore all linear invariants $f$ must be of the form 
$f=a_1 x_R+b_1 z_R$, from which we get the invariant $x_R$ and $z_R$ in (\ref{ccordinates}).

We then try to find {\it quadratic invariants} and therefore an $8 \times 8$ symmetric matrix $P$ so that
$f(\vec V, \vec W)=[\vec V^T, \vec W^T] P [\vec V^T, \vec W^T]^T$ and  
$$
[\vec V^T, \vec W^T] P [\vec V^T, \vec W^T]^T=[\vec V^T, \vec W^T] \begin{pmatrix}Ad_X^T & 0 \cr 0 & Ad_X^T \end{pmatrix}P   \begin{pmatrix}Ad_X & 0 \cr 0 & Ad_X \end{pmatrix} [\vec V^T, \vec W^T]^T, 
$$
for $X=X_z(\beta)$ and $X=X_{y}(\theta)$ as defined in (\ref{adXz}) and (\ref{adXy}) for every $\beta$ and $\theta$ (and for every $\vec V$ and $\vec W$). This leads to the condition 
$$
\begin{pmatrix} Ad_X & 0 \cr 
0 & Ad_X \end{pmatrix} P=P \begin{pmatrix} Ad_X & 0 \cr 
0 & Ad_X \end{pmatrix}.   
$$ 
From this, we find that the matrix $P$ must be of the form 
$$
P=\begin{pmatrix} e & 0 & 0 & 0& d & 0& 0 & 0\cr 
 0 & c & 0 & 0 &0 & g & 0 & 0 \cr 
 0 & 0 & c & 0 & 
 0 & 0 & g & 0 \cr 
 0 & 0 & 0 & c & 0 & 0 & 0 & g \cr 
 d & 0 & 0 & 0 & r & 0 & 0 & 0\cr 
 0 & g & 0 & 0 & 0 & h & 0 & 0 \cr 
 0 & 0 & g & 0 & 0 & 0 & h&  0 \cr 
 0 & 0 & 0 & g & 0 & 0 & 0 & h    \end{pmatrix}. 
$$
It follows that all quadratic invariants must be of the form 
$$
f=ex_R^2+2dx_R z_R + r z_R^2+c(x_I^2+y_R^2+y_I^2) + h (z_I^2+w_R^2+w_I^2)+ 2g(x_I z_I+w_R y_R+ w_I y_I). 
$$
Because of $x_R^2+x_I^2+y_R^2+y_I^2=z_R^2+z_I^2+w_R^2+w_I^2=1$, all terms can be  written in terms of the (linear) invariant $x_R$ and $z_R$ except the last one which we choose as the third coordinate in (\ref{ccordinates}). 

\subsection{Algorithm for control}\label{AlgoCon}

At the point $\pi(p) \in G_{reg}/SU(2)$, the tangent vectors 
$\frac{\partial}{\partial x^j},$ $j=1,2,3$ span $T_{\pi(p)} G_{reg}/SU(2)$, so that a general tangent vector at $\pi(p)$ can be written as $a_1\frac{\partial}{\partial x^1}+a_2 \frac{\partial}{\partial x^2}+ a_3 \frac{\partial}{\partial x^3}$. We calculate the matrix of the isomorphism $\pi_*|_{p}$  mapping 
the coordinates $\alpha_x, \alpha_y, \alpha_z,$ in $R_{p*} (\sigma, \gamma \sigma):=R_{p*}(\alpha_x(i \sigma_x, \gamma i \sigma_x) +\alpha_y (i \sigma_y, \gamma i \sigma_y)+ \alpha_z (i \sigma_z, \gamma i \sigma_z) ) \in R_{p*} {\cal P}$ to $(a_1,a_2,a_3)$, (cf. 
(\ref{diffsys})). Denote this matrix by $\Pi_{j,l}:=\Pi_{j,l}(p)$ with $j=1,2,3$ and $l=x,y,z$. 
We have 
$$
\Pi_{j,l}(p)= \pi_{*}|_p R_{p*} (i \sigma_l, i \gamma \sigma_l) x^j. 
$$
For the sake of illustration, let us calculate  $\Pi_{1,x}(p)$. This is given by (recall $p$ is defined in (\ref{puntop})) 
$$
\Pi_{1,x}(p):=\pi_{p*}R_{p*} (i \sigma_x, i \gamma \sigma_x) x^1=R_{p*} (i \sigma_x, i \gamma \sigma_x) (x^1 \circ \pi)=\frac{d}{dt}|_{t=0} x^1\circ \pi \left( e^{i \sigma_x t} \begin{pmatrix} x & y \cr -y^* & x^*\end{pmatrix}\, , \,e^{i \gamma \sigma_x t} \begin{pmatrix} z &  w \cr -w^* & w^*\end{pmatrix}  \right). 
$$
This simplifies because $x^1 \circ \pi $ does not depend on the second factor. Therefore the  $\Pi_{1,x}(p)$ entry is the derivative at $t=0$ of the real part of the $(1,1)$ entry of the matrix 
$$
e^{i \sigma_x t} \begin{pmatrix} x &  y \cr -y^* & x^*\end{pmatrix}  
= 
\begin{pmatrix} \cos(t) & i \sin(t) \cr i \sin(t) & \cos(t) \end{pmatrix} \begin{pmatrix} x &  y \cr -y^* & x^*\end{pmatrix}.
$$ 
This leads to  the result 
$$
\Pi_{1,x}(p)=-y_I. 
$$

The quantities 
\be{D1D2D3}
\Delta_1:=z_I y_R- x_I w_R, \qquad \Delta_2:=z_I y_I-x_Iw_I, \qquad \Delta_3:=w_Ry_I-w_Iy_R, 
\ee
appear routinely in calculations that follow.

Similar calculations to the ones above for $\Pi_{1,x}(p)$  lead to the full matrix $\Pi(p)$,  which is given by 
{
\be{PiofP}
\Pi(p):=
\begin{pmatrix} - y_I & -y_R & -x_I \cr 
- \gamma w_I & - \gamma w_R & - \gamma z_I \cr 
(\gamma -1) \Delta_1 +\gamma z_R y_I 
+w_I x_R 
& (1-\gamma) \Delta_2  
+ w_R x_R 
+ \gamma z_R y_R 
 &
(\gamma-1) \Delta_3 
+x_R z_I 
+ \gamma z_R x_I
\end{pmatrix}
\ee }
The determinant of this matrix is different from zero if and only if $p$ is in the regular part 
and it is another invariant under the action of $SU(2)$ on $SU(2) \times SU(2)$ (cf. Appendix B). It can be explicitly  computed as 
\be{determinante}
\det (\Pi (p))=\gamma(\gamma-1)(\Delta_1^2+\Delta_2^2+\Delta_3^2), 
\ee
which can be seen to be equal to zero if and only if  condition (\ref{commut34}) is verified. The invariant 
$\Delta:=\Delta_1^2+\Delta_2^2+\Delta_3^2$ can be expressed in terms of the (minimal) invariants $x_R$, $z_R$ and $x^3$ in (\ref{ccordinates}) as\footnote{This can be seen by expanding the left hand side using the definitions of $\Delta_{1,2,3}$ (\ref{D1D2D3}) and the right hand side using the definition of $x^3$ (\ref{ccordinates}), so that (\ref{delt}) reduces to $y_I^2w_R^2+w_I^2y_R^2+y_R^2 z_I^2+x_I^2 w_R^2+x_I^2 w_I^2+y_I^2 z_I^2=(1-x_R^2)(1-z_R^2)-z_I^2x_I^2-y_R^2 w_R^2-y_I^2 w_I^2$, and writing $(1-x_R^2)=x_I^2+y_R^2+y_I^2$ and  $(1-z_R^2)=z_I^2+w_R^2+w_I^2$, we obtain an identity.}
\be{delt}
\Delta=\Delta_1^2+\Delta_2^2+\Delta_3^2=(1-x_R^2)(1-z_R^2)-(x^3)^2. 
\ee

When we design a control law, the components $a_1,a_2,a_3$ of the tangent vector  at every time $t$ in the tangent space at $\pi(p(t))$ are given by the derivatives $\dot x^1$, $\dot x^2$, $\dot x^3$ of the desired trajectory in the quotient space. The corresponding components,  $\alpha_x$, $\alpha_y$ and $\alpha_z$, of the tangent vector in $R_{p(t)*}{\cal P}$ give  the appropriate control functions $(u_x, u_y, u_z)$. The matrix $\Pi(p)$ in (\ref{PiofP}) gives the map from the control to trajectories. Since we want to specify trajectories and compute the corresponding controls, we need the inverse of the matrix $\Pi(p)$ (cf. (\ref{deficontr})). This is found from (\ref{PiofP}) to be 
\be{Piinverse}
\det(\Pi(p)) \Pi^{-1}(p):=
\begin{pmatrix}
\gamma(\gamma -1)(-w_R \Delta_3 - z_I \Delta_2) + \gamma^2z_R \Delta_1 &
(\gamma -1)(x_I \Delta_2 + y_R \Delta_3)+ x_R \Delta_1 & \gamma \Delta_1 \cr 
\gamma (\gamma -1)(w_I \Delta_3 - z_I \Delta_1)- \gamma^2 z_R \Delta_2 & 
(\gamma -1) (x_I \Delta_1 - y_I \Delta_3) - x_R \Delta_2 & 
- \gamma \Delta_2 \cr 
\gamma ( \gamma-1) ( w_I \Delta_2 + w_R \Delta_1) + \gamma^2 z_R \Delta_3 
& 
-(\gamma-1)(y_I \Delta_2 +y_R \Delta_1) + x_R \Delta_3 
& 
\gamma \Delta_3
\end{pmatrix}. 
\ee

We remark that $\Pi^{-1}(p)$ is not defined if we are in the singular part of the space $G=SU(2) \times SU(2)$ as the determinant of $\Pi$ is zero there. This is in particular true at the beginning as the initial point $p \in SU(2) \times SU(2)$ is the identity. In order  to follow a  prescribed trajectory in the quotient space $G_{reg}/SU(2)$, we need therefore to apply a preliminary control to drive the state to an arbitrary point in $G_{reg}$ and after that we 
shall apply the control  corresponding to a prescribed trajectory in the quotient space.

The preliminary control in an interval $[0,T_1]$ to move the state from the singular part of the quotient space has to involve at least two different directions in the tangent space. In other terms, if we use $\sigma(t)=u(t) \sigma$ for some function $u=u(t)$ and a constant matrix $\sigma \in \mathfrak{su}(2)$ we remain in the singular part. To see this, notice that if 
$d:=\int_0^{T_1} u(s)ds$, then the solution of  (\ref{reple}) will be $(U_f, V_f)=(e^{d \sigma},  e^{\gamma d \sigma})$, a pair that satisfies the condition (\ref{commut34}). Therefore the simplest control strategy of moving in one direction only will not work if we want to move the state from the singular part. Furthermore, we want $u_x(0)=u_y(0)=u_z(0)=0$ and $u_x(T_1)=u_y(T_1)=u_z(T_1)=0$ to avoid discontinuities at the initial time $t=0$ and at the time of concatenation with the second portion  of the control. We propose to prescribe a trajectory 
for $U=U(t)$ in  (\ref{reple}) and, from that trajectory, to  derive  the control to be used in the equation for $V=V(t)$ in (\ref{reple}). We choose a smooth function $\delta:=\delta(t)$ such that 
$\delta(0)=0$ and $\delta(T_1)=\delta_0 \not=0 $, and $\dot \delta(0)=\dot \delta(T_1)=0$. 
We also choose a smooth function $\epsilon:=\epsilon(t)$, with $\epsilon(0)=\epsilon_0 \not=0 $ and $\epsilon(T_1)=0$, and $\dot \epsilon(0)=\dot \epsilon(T_1)=0$. We choose for $U=U(t)$ in (\ref{reple}) 
\be{Uoft}
U(t)=\begin{pmatrix} \cos(\delta(t)) & e^{i \epsilon(t)} \sin(\delta(t)) \cr 
-e^{-i\epsilon(t)} \sin(\delta(t)) & \cos(\delta(t)) \end{pmatrix}, 
\ee
which at time $T_1$ gives 
\be{Poil}
U(T_1)=\begin{pmatrix} \cos(\delta_0) &  \sin(\delta_0) \cr 
-\sin(\delta_0)   & \cos(\delta_0) \end{pmatrix}. 
\ee
The corresponding control $\sigma$ is $\sigma(t)=\dot U U^\dagger$,  which is 
\be{controlSS}
\sigma(t)=\begin{pmatrix} i \dot \epsilon \sin^2(\delta) & \dot \delta e^{i \epsilon}+ \frac{i}{2} \, \dot \epsilon  \,
{\sin(2 \delta)}\, e^{i\epsilon}  \cr 
- \dot \delta e^{-i \epsilon}+ \frac{i}{2} \dot \epsilon \,  {\sin(2 \delta)} \, e^{-i\epsilon} & -i \dot \epsilon \sin^2(\delta) \end{pmatrix}.
\ee
Placing this in the second equation of (\ref{reple}) and integrating numerically we obtain the values for $V(T_1)$,  the second component of $(U,V)$, and therefore  the values of $(z_R,z_I,w_R,w_I)$. Using these values and the expression for $(x_R,x_I,z_R,z_I)$ in (\ref{Poil}), and using the formula  for $\Delta$  given in  
(\ref{delt}), we obtain 
\be{accor5}
\Delta=\Delta_1^2+\Delta_2^2+\Delta_3^2=\sin^2(\delta_0)(1-z_R^2(T_1)-w_R^2(T_1))=\sin^2(\delta_0)(z^2_I(T_1)+w^2_I(T_1)),
\ee
which has to be different from zero in order for the state to be in the regular part.

The second portion  of the control depends on the trajectory followed, $(x^1,x^2,x^3)=(x^1(t),x^2(t),x^3(t))$, and it is obtained by multiplying by $\Pi^{-1}$ in (\ref{Piinverse}) $(\dot x^1,\dot x^2, \dot x^3)$. The trajectory $(x^1,x^2,x^3)$ is almost completely arbitrary. However it has to satisfy certain conditions which we now discuss.  Let us denote the interval where the second part of the control is used by $[0,T_2]$. The initial condition $(x^1(0),x^2(0),x^3(0))$ has to agree with the one given by the previous interval of  control. The final condition  $(x^1(T_2),x^2(T_2),x^3(T_2))$ has to agree with the orbit of the  desired final condition. Moreover, care has to be taken to make sure that the trajectory is such that $\Delta$  in (\ref{delt}) is never zero because this would create a singularity in $\Pi^{-1}(p)$. Furthermore 
we need  $\dot x^1(0)=\dot x^2(0)=\dot x^3(0)=0$, which gives $\sigma(0)=0$, 
 to ensure continuity  with the  control in the previous interval, and we also choose $\dot x^1(T_2)=\dot x^2(T_2)=\dot x^3(T_2)=0$ to ensure that the control is switched off at the end of the procedure. Finally, the functions $(x^1, x^2, x^3)$ have to be representative of a possible trajectory for special unitary matrices. This means that,  with $\vec V:=(x_R, x_I, y_R, y_I)^T$ and $\vec W:=(z_R, z_I, w_R, w_I)^T$, 
 $\|\vec V (t)\|^2=\|\vec W(t)\|^2=1$, at every $t$. Therefore $|x_R(t)| < 1$ at every $t$, $|z_R(t)| < 1$, at every $t$ (to avoid singularity), and from the Schwartz inequality $|\vec V \cdot \vec W |\leq 1$ we also must have $|x_R z_R + x^3|\leq 1$ and therefore 
\be{condx3} 
 -1-x_R z_R \leq  x^3 \leq  1 - x_R z_R. 
\ee

Once the functions $(\dot x^1, \dot x^2,\dot x^3)$ are chosen, the system to integrate numerically is (\ref{reple}) with $(u_x,u_y,u_z)$ given by $\Pi^{-1}(p)(\dot x^1, \dot x^2,\dot x^3)^T$. By deriving $(u_x,u_y,u_z)$ using the explicit expression of $\Pi^{-1}$ given in (\ref{Piinverse}) and replacing into (\ref{reple}), it is possible to obtain a  simplified system of differential equations for $(x_R,x_I,...,z_R,z_I,...,w_I)$ without implementing the preliminary step of calculating the control. We found this system to be more stable in numerical integration with MATLAB and report it in Appendix A for future use.

\subsection{Numerical example: Driving to two different Hadamard  gates} 

We now apply the above technique to  a specific numerical example: The problem is  to drive 
the system (\ref{Hamilto}) so that the first spin performs the Hadamard-type  gate 
\be{Hada1}
H_1:=\begin{pmatrix} \frac{1}{\sqrt{2}} & \frac{1}{\sqrt{2}} \cr 
\frac{-1}{\sqrt{2}} & \frac{1}{\sqrt{2}}\end{pmatrix}
\ee
and the second spin performs the Hadamard gate
\be{Hada2}
H_2:=\begin{pmatrix} \frac{1}{\sqrt{2}} & \frac{-i}{\sqrt{2}} \cr 
\frac{-i}{\sqrt{2}} & \frac{1}{\sqrt{2}}\end{pmatrix}. 
\ee
We want to  drive system (\ref{reple}) to $(U_f, V_f)=(H_1, H_2)$. The orbit of the desired final condition is characterized by the invariant coordinates 
\be{finalorbit}
x^1=x_R=\frac{1}{\sqrt{2}}, \qquad x^2=z_R=\frac{1}{\sqrt{2}}, \qquad x^3=x_I z_I+ y_R w_R+ y_I w_I=0. 
\ee
We take  a physical value for the ratio $\gamma$ between the two gyromagnetic ratios. In particular we will choose 
  $\gamma\approx \frac{1}{0.2514}$ which corresponds to the Hydrogen-Carbon ($^{1} H-^{13} C$) system also considered in  \cite{Xinhua}. 

We first consider the control that moves the state away from the singular part, in a time  interval $[0,1]$. We choose 
$\sigma$ in (\ref{controlSS}) with  the functions 
$\delta$ and $\epsilon$ as follows:  
\be{deltaandepsilon}
\delta=6 \delta_0 \left( \frac{t^2}{2} -\frac{t^3}{3} \right), \qquad \epsilon=\epsilon_0 + 6 \epsilon_0
\left( \frac{t^3}{3} -\frac{t^2}{2} \right). 
\ee
With these functions $\delta$ and $\epsilon$, $\sigma$  satisfies all the requirements described above. 
From (\ref{deltaandepsilon}) and (\ref{controlSS}) we obtain the controls $u_x,u_y,u_z$ which replaced into (\ref{reple}) give the dynamics in the interval $[0,T_1]=[0,1]$. Numerical integration with the values of the parameters $\delta_0=0.5$ and $\epsilon_0=1$, gives the following conditions at time $T_1=1$ (cf. (\ref{Poil}) 
\be{condizioni}
U(1)=\begin{pmatrix} \cos(0.5) & \sin(0.5) \cr 
-\sin(0.5) & \cos(0.5) \end{pmatrix}, \qquad V(1) \approx \begin{pmatrix} -0.3472+i0.7769 & -0.5123 -i 0.1157\cr 
0.5123-i0.1157  &  -0.3472-i0.7769 \end{pmatrix}. 
\ee
The value of $\Delta$ is, according to (\ref{accor5}), 
$\Delta \approx \sin^2(0.5)\left( (0.7769 )^2+(0.1157)^2 \right) \not= 0$, as desired.

\begin{figure}[htb]
\centering
\includegraphics[width=\textwidth, height=0.4\textheight]{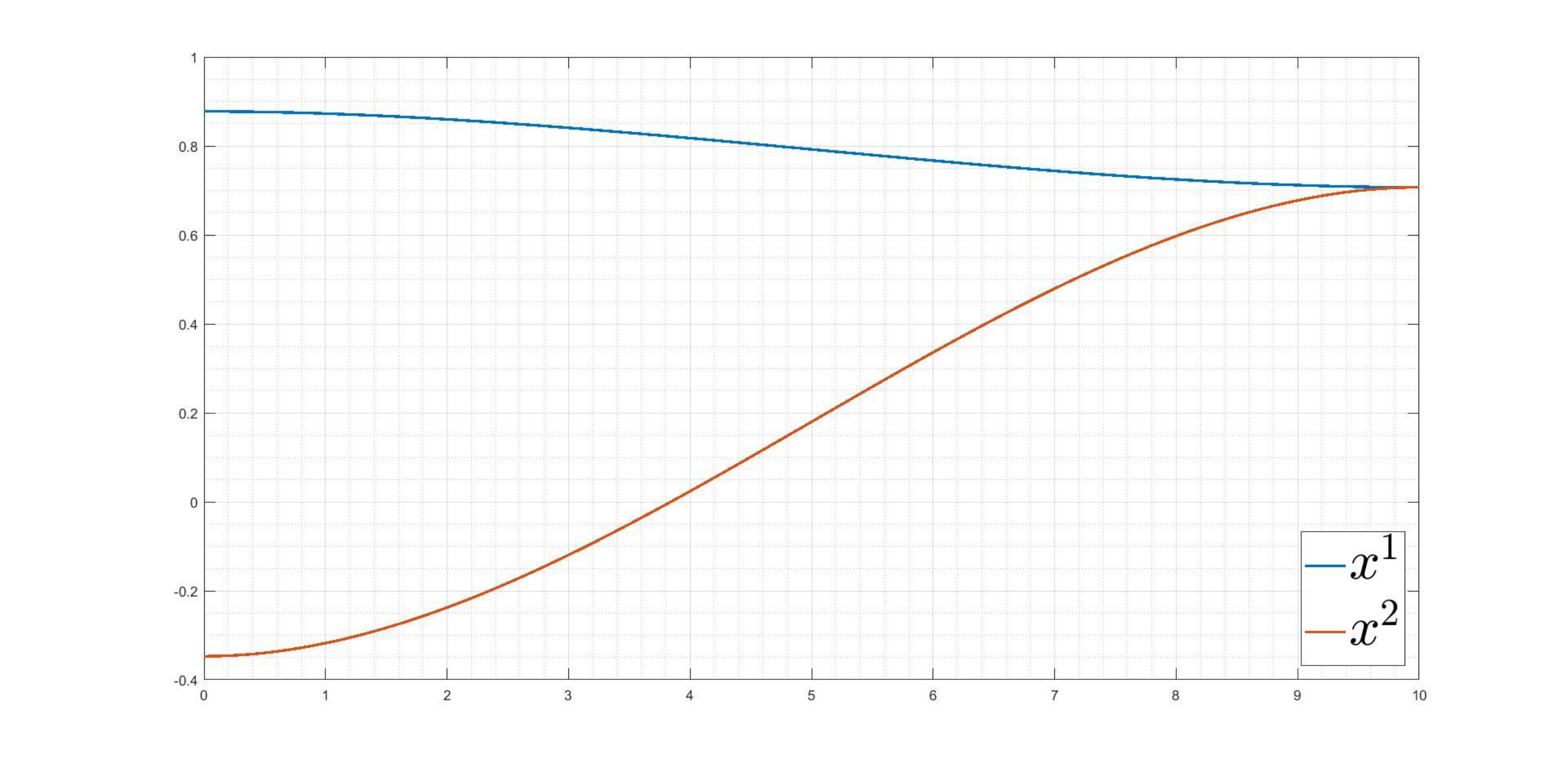}
\vspace*{-16mm}
\caption{Prescribed $x^1$ and $x^2$ variables in the (second) interval $[0,10]$.}
\label{Fig1}
\end{figure}

\begin{figure}[htb]
\centering
\includegraphics[width=\textwidth, height=.4\textheight]{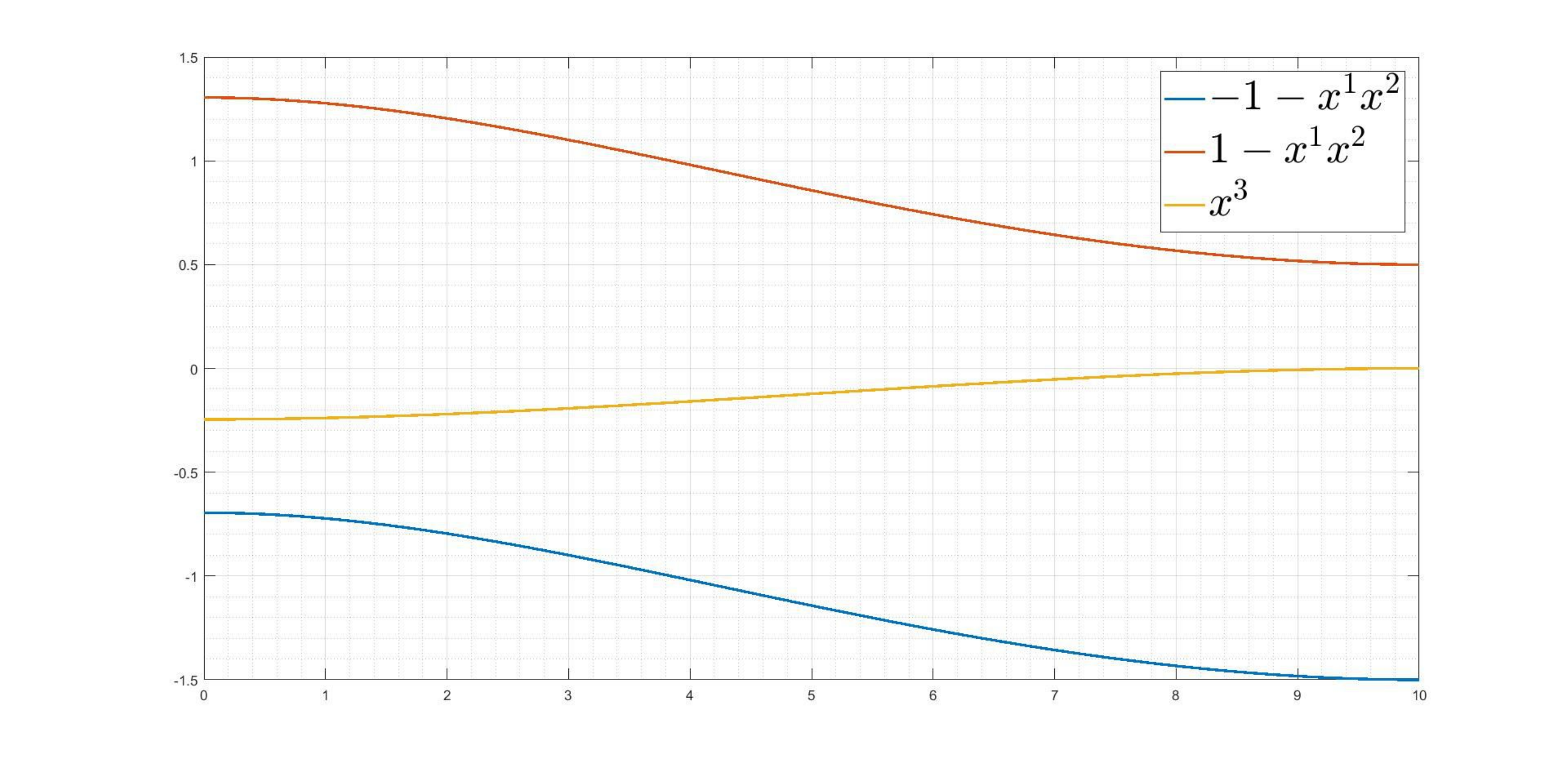}
\vspace*{-16mm}
\caption{$x^3=x^3(t)$ vs $-1-x^1(t)x^2(t)$ and $ 1-x^1(t)x^2(t)$ in the (second) interval $[0,10]$.}
\label{Fig2}
\end{figure}

\begin{figure}[htb]
\centering
\includegraphics[width=\textwidth, height=.4\textheight]{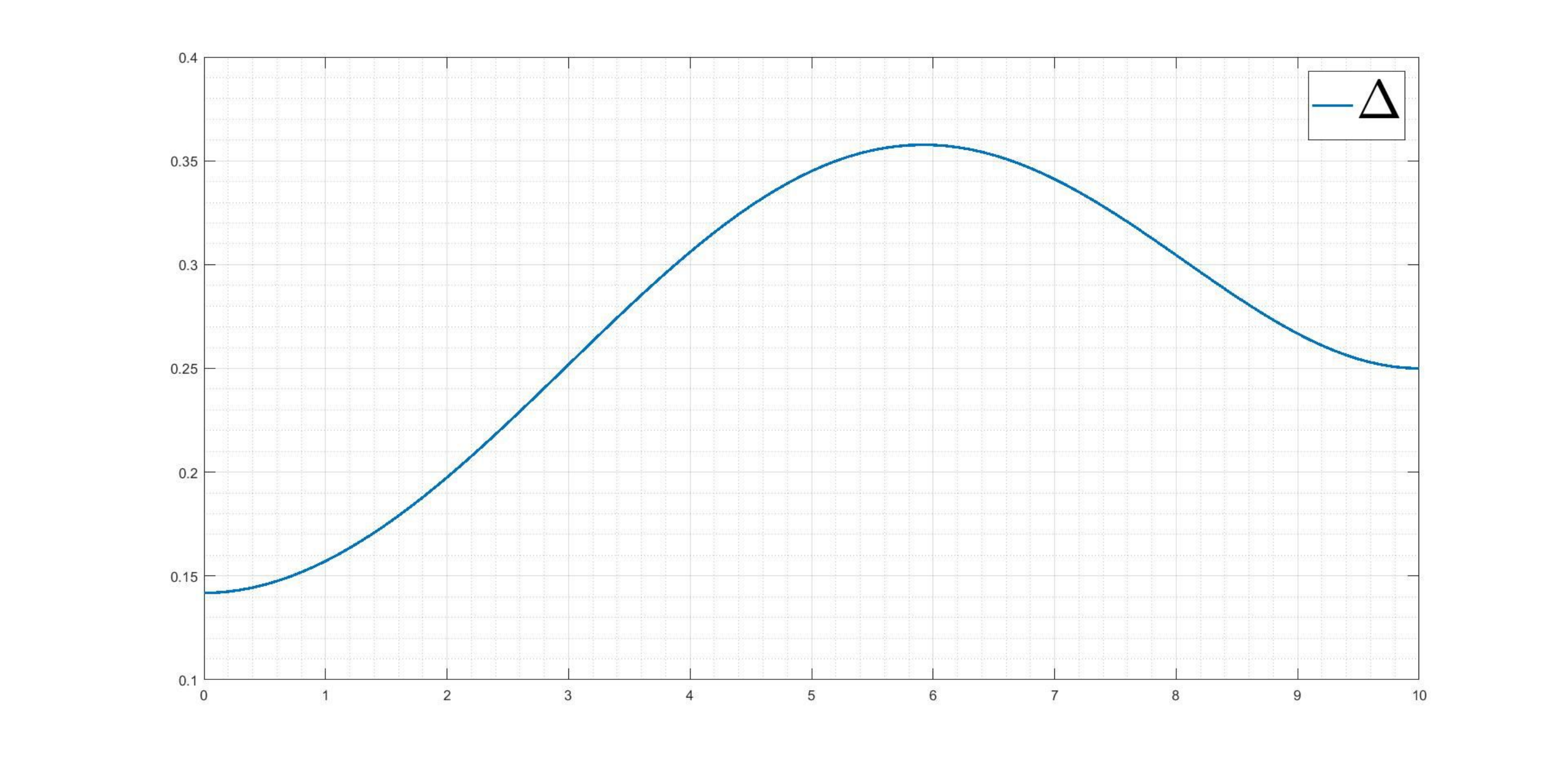}
\vspace*{-16mm}
\caption{$\Delta=\Delta(t)=\Delta_1^2(t)+\Delta_2^2(t)+\Delta_3^2(t)$ in the (second) interval $[0,10]$.}
\label{Fig3}
\end{figure}

The values of the variables to be used as initial conditions in the integration in the subsequent interval of the procedure are 
$
x_R(1)=x^1(1)=\cos(0.5), \quad x_I(1)=0, \quad y_R(1)=\sin(0.5), \quad y_I(1)=0, \quad 
z_R(1)=x^2(1)=-0.3472 , \quad z_I(1)=0.7769 , \quad w_R(1)=-0.5123, \quad w_I(1)=-0.1157, 
$ 
and $x^3(1)=x_I(1) z_I(1)+y_R(1) w_R(1)+y_I(1) w_I(1)=\sin(0.5) \times (-0.5123)\approx -0.2456.$ For the subsequent 
 interval $[0,T_2]$ we choose the trajectory $x^1(t),x^2(t),x^3(t)$ in the quotient space as follows: 
$T_2=10$ and the trajectory in the interval $[0,T_2]$ is 
\be{x1t}
x^1(t)=-\frac{1}{500} \left( \frac{1}{\sqrt{2}}-\cos(0.5)\right)t^3+\frac{3}{100} \left( \frac{1}{\sqrt{2}} - \cos(0.5) \right)t^2+ \cos(0.1);
\ee
\be{x2t}
x^2(t)=-\frac{1}{500} \left( \frac{1}{\sqrt{2}}+0.3472 \right) t^3+\frac{3}{100} \left( \frac{1}{\sqrt{2}}+
0.3472 \right)t^2-0.3472; 
\ee  
\be{x3t}
x^3(t):=\frac{-0.2456}{500} t^3-\frac{3 \times (-0.2456)}{100} t^2-0.2456,  
\ee
which are easily seen to satisfy the conditions at the endpoints. Moreover by plotting $x^1$ and $x^2$ we see that 
$|x^1(t)|\leq 1$ and $|x^2(t)|\leq 1$ for every $t \in [0,10]$ (Figure \ref{Fig1}). 
By plotting $x^3$ vs $1-x^1 x^2$ and $-1-x^1x^2$ (Figure \ref{Fig2}) we find that $-1-x^1 x^2(t) \leq x^3(t) \leq 1-x^1 x^2(t)$ for every $t \in [0,10]$ as required from condition (\ref{condx3}). 
By plotting $\Delta=\Delta(t)$ in $[0,10]$ we know that $\Delta(t) \not=0$ for every $t \in [0,10]$ (Figure \ref{Fig3}). Therefore the whole trajectory is in the regular part. 

\begin{figure}[htb]
\centering
\includegraphics[width=\textwidth, height=.4\textheight]{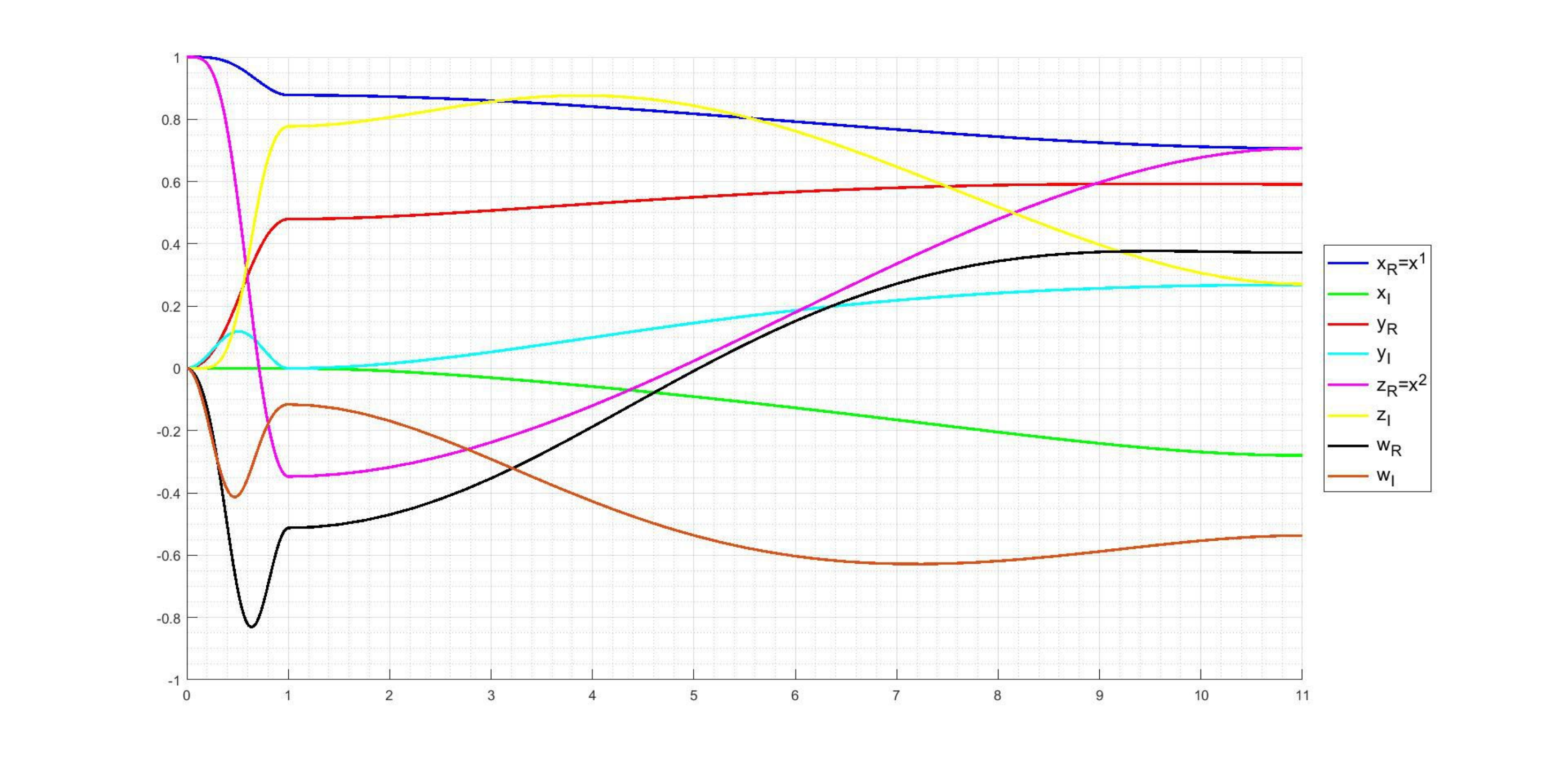}
\vspace*{-16mm}
\caption{Trajectory under the preliminary control in the total interval $[0,11]$.}
\label{Fig4}
\end{figure}

The {\it full} trajectory, in the union  of the two intervals, and with the concatenation 
of the two controls is depicted in Figure \ref{Fig4}. Let us denote the full control by $\hat \sigma=\hat \sigma(t)=
u_x i\sigma_x+ u_y i \sigma_y+ u_z i \sigma_z$. The final condition $(\hat U_f, \hat V_f)$ is given by 
\be{Finalprel}
\hat U_f=\begin{pmatrix}  0.7071-0.2795i  & 0.5913+0.2685i \cr 
-0.5913+0.2685i & 0.7071+0.2795i \end{pmatrix}, 
\qquad \hat V_f=\begin{pmatrix}  0.7071+0.2708i  &  0.3718-0.5369i \cr 
 -0.3718-0.5369i & 0.7071-0.2708i \end{pmatrix}.  
\ee 

\begin{figure}[htb]
\centering
\includegraphics[width=\textwidth, height=.4\textheight]{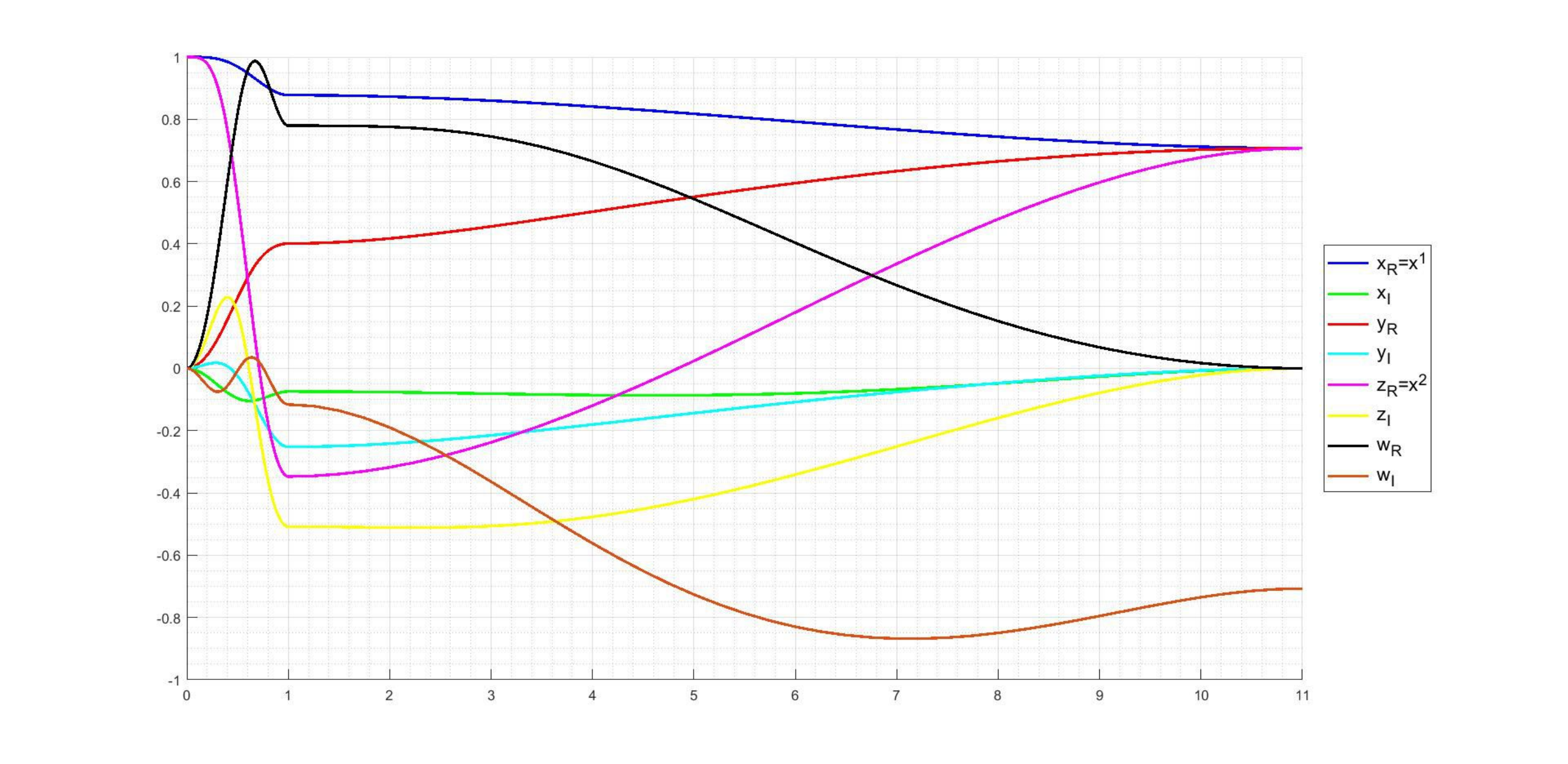}
\vspace*{-16mm}
\caption{Trajectory under control algorithm leading to the desired final condition (\ref{Hada1}) and (\ref{Hada2}).}
\label{Fig5}
\end{figure}

This condition, as expected, is in the same orbit as the desired final condition $(H_1,H_2)$  in (\ref{Hada1}), (\ref{Hada2}), that is, there exists a matrix $K \in SU(2)$ such that
$K\hat U_fK^\dagger=H_1$ and $K\hat V_fK^\dagger=H_2$. The matrix $K$ solving these equations is found to be 
$$
K=\begin{pmatrix}  0.1485-0.2460i &  -0.2444+0.9260i \cr 
 0.2444+0.9260i & 0.1485+0.2460i\end{pmatrix}. 
$$
In particular, to find $K$ one can diagonalize $\hat U_f$ and $H_1$, i.e., $\hat U_f=P \Lambda P^\dagger$ and $H_1=R\Lambda R^\dagger$ for a diagonal matrix $\Lambda$, so that, from $KP\Lambda P^\dagger K^\dagger=R \Lambda R^\dagger$,  we find that $R^\dagger K P=D$, for $D$, a diagonal matrix. This matrix is found by solving $DP^\dagger \hat V_f P=R^ \dagger H_2 R D$. The control $K \hat \sigma K^\dagger$ steers then to the desired final condition. The resulting trajectory leading to the  desired final condition (\ref{Hada1}), (\ref{Hada2}) is given in Figure \ref{Fig5}.   

\section{Appendix A: System of ODE's for the simultaneous control of two quantum bits in Subsection \ref{AlgoCon}}

We derived the system of ODE's (\ref{reple}) with $(u_x,u_y,u_z)$ obtained  from  $(u_x,u_y,u_z)^T=\Pi^{-1}(p)(\dot x^1, \dot x^2, \dot x^3)^T$ with $\Pi^{-1}$ given in (\ref{Piinverse}). We define in the following $\Delta:=\Delta_1^2+\Delta_2^2+\Delta_3^2$, $Z:=x^3:=x_Iz_I+y_R w_R+y_I w_I$.  Simplifications are obtained using the following two relations which are directly verified. 
$$
y_I\Delta_1-y_R\Delta_2+x_I\Delta_3=0,
$$
$$
w_I\Delta_1-w_R\Delta_2+z_I\Delta_3=0.
$$
The system becomes with $a:=\dot x^1$, $b:=\dot x^2$, $c:=\dot x^3$ 
\small{
\begin{flalign}
& \dot x_R=a \nonumber \\
& \gamma(\gamma -1) \Delta \dot x_I=(1-\gamma) \Delta_3 b+(\gamma^2z_Ra +\gamma c+ \gamma b x_R)(x_R \Delta_3-y_I \Delta_2-y_R\Delta_1)+ \gamma (\gamma -1) a ( \Delta_3 Z+ x_R(w_I \Delta_2+ w_R \Delta_1)) \nonumber \\
& \gamma(\gamma -1) \Delta \dot y_R=(\gamma-1) \Delta_2 b + (\gamma^2z_Ra +\gamma c+ \gamma b x_R)(x_I \Delta_1-x_R \Delta_2-y_I \Delta_3)+ \gamma (\gamma -1) a ( - \Delta_2 Z+ x_R(w_I \Delta_3- z_I \Delta_1)) \nonumber \\
& \gamma(\gamma -1) \Delta \dot y_I=(1-\gamma) \Delta_1 b+ (\gamma^2z_Ra +\gamma c+ \gamma b x_R)(x_R \Delta_1+ x_I \Delta_2+ y_R \Delta_3)+\gamma (\gamma -1) a ( \Delta_1 Z- x_R(w_R \Delta_3+ z_I \Delta_2)) \nonumber \\
&\dot z_R=b \nonumber \\
&(\gamma -1) \Delta \dot z_I=\gamma(\gamma -1) \Delta_3 a+(\gamma z_R a+ \gamma c+ x_R b)(z_R \Delta_3-w_R \Delta_1-w_I \Delta_2) +(1-\gamma)b( \Delta_3 Z+z_R(y_R\Delta_1+y_I \Delta_2)) \nonumber \\
&(\gamma -1) \Delta \dot w_R=\gamma(1-\gamma) \Delta_2 a+(\gamma z_R a+ \gamma c+ x_R b)(z_I \Delta_1-\Delta_3 w_I-z_R\Delta_2)+ (\gamma-1)b(\Delta_2 Z+z_R(x_I \Delta_1-y_I \Delta_3)) \nonumber \\
&(\gamma -1) \Delta \dot w_I=\gamma(\gamma-1) \Delta_1 a+(\gamma z_R a+ \gamma c+ x_R b)(\Delta_1 z_R+z_I \Delta_2+w_R \Delta_3) + (\gamma -1)b ( - \Delta_1 Z + z_R(y_R \Delta_3+ x_I \Delta_2)) \nonumber 
\end{flalign}}

\section{Appendix B : Invariance of the determinant of $\pi_*$} 

Let ${\cal L}$ be a semisimple Lie algebra with decomposition ${\cal L}={\cal K} \oplus {\cal P}$ and 
$[{\cal K},{\cal K}] \subseteq {\cal K}$ and  $[{\cal K},{\cal P}] \subseteq {\cal P}$, and consider the conjugacy action of $e^{\cal K}$ on $e^{\cal L}$. Consider the natural projection $\pi: \, e^{\cal L} \rightarrow e^{\cal L}/e^{\cal K}$ and $p \in e^{\cal L} $ a regular point so that at $p$, $\pi_*$ is an isomorphism $\pi_* \, : \, R_{p*}{\cal P}   \rightarrow T_{\pi(p)} e^{\cal L}/ e^{\cal L}$. Given  bases in 
$R_{p*}{\cal P}$ and  $T_{\pi(p)} e^{\cal L}/ e^{\cal L}$ the matrix $\Pi=\Pi(p)$ representing $\pi_*$ has  determinant which is invariant  under the action of $e^{\cal K}$, i.e., for every $k \in e^{\cal K}$
$$
\det {\Pi(p)}=\det {\Pi(kpk^{-1})}.
$$

\bpr 
Let $\{B_1,...,B_m\}$ be a basis of ${\cal P}$ so that $\{R_{p*}B_1,...,R_{p*}B_m\}$ is a basis of $R_{p*{\cal P}}$. Let $x^1,..., x^m$ be a set of coordinates for $e^{\cal L}_{reg}/e^{\cal K}$ in a neighborhood of $\pi(p)$. The $ m \times m$ matrix $\Pi$ has entries 
$$
\Pi_{j,l}(p)=(\pi_* R_{p*}B_l) x^j, 
$$
and it maps a vector $[\alpha^1,...,\alpha^m]^T$ representing a tangent vector $V _1\in R_{p*}{\cal P}$, 
i.e., $V_1:=\sum_{l=1}^m \alpha^l R_{p*} B_l$ to a vector $[r^1,...,r^m]^T$ representing  
a tangent vector $ V_2 \in T_{\pi(p)} e^{\cal L}_{reg}/ e^{\cal K}$, i. e., $V_2=\sum_{j=1}^m r^j \frac{\partial}{\partial x^j}$. 


Let $P \in {\cal P}$ with $P:=\sum_{l=1}^m \alpha^l B_l$. Then, with $k \in e^{\cal K}$, we obtain\footnote{This follows from the definitions. For any function $f$,  we have $\pi_* R_{p* P} f=R_{p*} P (f \circ \pi)=
\frac{d}{dt}|_{t=0} f \circ \pi (e^{Pt} p)=\frac{d}{dt}|_{t=0} f \circ \pi (k e^{Pt} p k^{-1})= 
\frac{d}{dt}|_{t=0} f \circ \pi (e^{kPk^{-1}t} k p k^{-1})=\pi_* R_{kpk^{-1}} kPk^{-1}$.}
$$
\pi_*R_{p*}P=\pi_* R_{kpk^{-1}} kPk^{-1}.
$$
Therefore we have 
$$
\Pi_{jl}(p)=(\pi_* R_{p*} B_l) x^j= (\pi_* R_{kpk^{-1}*} k B_l k^{-1}) x^j=
$$
$$
(R_{kpk^{-1}*} k B_l k^{-1}) x^{j} \circ \pi = \frac{d}{dt}|_{t=0} x^j \circ \pi ( e^{kB_l k^{-1} t} k p k^{-1}).  
$$
Write $k B_l k^{-1}$ as $k B_l k^{-1}=\sum_{s=1}^m a_l^s B_s$, 
for an orthogonal matrix\footnote{The fact that the matrix  $\{a_s^l\}$, representing the adjoint action, is orthogonal is a consequence of the fact that the inner product, which is the Killing form on ${\cal P}$ is bi-invariant, and therefore it is not changed by the adjoint action.} $a_s^l$. Therefore we have 
$$
\Pi_{jl}(p)=\frac{d}{dt}|_{t=0} x^j \circ \pi ( e^{kB_l k^{-1} t} k p k^{-1})=
\frac{d}{dt}|_{t=0} x^j \circ \pi ( e^{\sum_{s=1}^m a_l^s B_s  t} k p k^{-1})=
$$
$$
\sum_{s=1}^m a_{l}^s \frac{d}{dt}|_{t=0}(x^j \circ \pi ( e^{B_s t} kpk^{-1})=
\sum_{s=1}^m a_l^s \Pi_{j,s}(kpk^{-1}). 
$$
Therefore there exists an orthogonal matrix $A=A(k)$ so that 
$$
\Pi(p)=\Pi(kpk^{-1}) A(k). 
$$
Taking the determinant of this relation and using $\det(A(k))=1$\footnote{Note that $\det(A(k))\neq -1$, since $A:e^{\cal K}\to O(m,\mathbb{R})$; $A\mapsto A(k)$ is continuous and $\det(A(k))=\det({\bf 1}_m)=1$.}, we obtain $\det(\Pi(p))= \det(\Pi(kpk^{-1}))$ as desired. 
\epr

\end{document}